\documentclass[12pt]{article}
\usepackage{graphicx}
\usepackage{amsmath}
\usepackage{bm}% bold math
\def\Vec#1{\mbox{\boldmath $#1$}}

\def\itmb{\begin{itemize}}
\def\itme{\end{itemize}}
\def\enmb{\begin{enumerate}}
\def\enme{\end{enumerate}}
\def\eqnb{\begin{equation}}
\def\eqne{\end{equation}}

\def\NPB{{Nucl. Phys.} { B}}
\def\PLB{{Phys. Lett.} B}
\def\PRL{Phys. Rev. Lett.}
\def\PRD{{Phys. Rev.} D}
\def\PRC{{Phys. Rev.} C}

\title{Axial anomaly and the triality symmetry\\ of leptons and hadrons }

\author{Sadataka Furui \\
  Faculty of Science and Engineering, Teikyo University\\
1-1 Toyosatodai, Utsunomiya, 320-8551 Japan {\thanks
{\textit{E-mail address:} furui@umb.teikyo-u.ac.jp}}
}

\begin{document}
\maketitle
\begin{abstract}%
We apply the supersymmetric model of \'E. Cartan to the pseudoscalar meson decay into two photons, $\pi_0\to\gamma\gamma$, $\eta\to\gamma\gamma$ and $\eta'\to\gamma\gamma$.
In the book of \'E. Cartan published in 1966, Dirac spinors ${^t(}A,B)$ and ${^t(}C,D)$ and vector fields $E$ and $E'$ were introduced and five supersymmetric transformations $G_{23}, G_{12}, G_{13}, G_{123}$ and $G_{132}$ were considered. 

The Pauli spinor is treated as a quaternion and the Dirac spinor is treated as an octonion. In the pseudoscalar meson decay, when the two final vector fields belong to the same group ($EE$ or $E'E'$), we call the diagram rescattering diagram.  When they belong to different groups ($EE'$), the diagram is called twisted diagram.  

Assuming the triality selection rules of octonions, dark matter is interpreted as matter emitting photons  in a different triality sector than that of electromagnetic probes in our world.

\end{abstract}
%\keywords{Triality, Quaternion, Octonion}

\newpage
\section{Introduction}
According to  Hawking and Mlodinow, physical theories are assembly of mathematical models or assembly of rules that connect elements of models and observables\cite{HM11}.   In QED, complex numbers and quaternion are used and in QCD also the same number system is used.  A quaternion operates on a two-component spinor. Pauli spinor is a two-component spinor, but Dirac spinor is a four component spinor, and there is an octonion that operates on a four component spinor, which has the triality symmetry.   

In the low-energy world there are two kinds of Dirac spinors, leptons $e, \mu$ and $\tau$, and quarks $u, d$ and $s$.  Quarks have three-color degrees of freedom. When color degrees of freedom is large, the quark loop degrees of freedom $U(1)$-anomaly due to difference of masses of $e,\mu$ and $\tau$ is suppressed, but in the infrared there appears quark condensates $\langle 0|\bar q q|\rangle\ne 0$, and $U(1)$ symmetry is broken and there appears $\eta'$ meson in addition to $\pi^\pm, \pi^0$, $K^\pm, K^0,\bar K_0$ and $\eta$, that plays essential roles\cite{GL84}.  

The anomaly was accounted for in the effective theory by the Wess-Zumino-Witten term\cite{WZ71, Witten83} and the effective Lagrangian in the presence of external source 
\begin{eqnarray}
\mathcal L_{QCD}&=&\mathcal L^0_{QCD}+\bar q\gamma_\mu(v^\mu+\gamma_5 a^\mu)q-\bar q(s-i\gamma_5 p)q\nonumber\\
 &&-\frac{g^2}{16\pi^2}\theta\, \epsilon^{\alpha\beta\lambda\mu}tr_c(G_{\alpha\beta}{G}_{\mu\nu})
\end{eqnarray}
with the vacuum angle $\theta$ was introduced in\cite{KL00}, and studied for a study of the mixing of $\eta$ and $\eta'$\cite{BW01,BB01,BN03, KK06, BCLRT07}.

The singlet axial current is normalized by introducing a number of flavors $N_f$ as
\begin{equation}
{A^0}_\mu=\frac{1}{\sqrt{N_f}}\bar q\gamma_\mu \gamma_5q
\end{equation}
and the divergence
\begin{equation}
\partial^\mu {A^0}_{\mu}=\sqrt{2N_f}\frac{1}{16\pi^2}tr_c \epsilon^{\alpha\beta\mu\nu}G_{\alpha\beta}{G}_{\mu\nu}= \sqrt{2N_f}\omega
\end{equation}

The $\eta'$ field defined in \cite{KL00} is
\[
\left\langle 0|{A^0}_\mu|\eta'\right\rangle=ip_\mu F_0, \qquad \left\langle 0|\omega|\eta'\right\rangle=\sqrt{2N_f}\frac{\tau}{F_0}
\]
where the mass of $\eta'$ and the parameter $\tau$ are related by
\begin{equation}
{M_{\eta'}}^2=\frac{{2 N_f}\tau}{F_0^2}
\end{equation}

The triangle diagram yields the electromagnetic anomaly of the axial quark charge current
\[
\partial_\mu j_\mu^{5\lambda}=-\frac{e^2}{16\pi^2}\epsilon^{\alpha\beta\mu\nu}F_{\alpha\beta}F_{\mu\nu}\cdot tr[\tau^\lambda Q^2]
\] 
where $F_{\mu\nu}$ is the electromagnetic field strength, 
\[
Q=\left(\begin{array}{cc}\frac{2}{3}&0\\
                                 0&-\frac{1}{3}\end{array}\right)
\]
is the quark electric charge and $tr$ is over colors and flavors\cite{PS95}.

The Nambu-Goldstone boson $\pi$'s decay into two gamma rays is described by divergence of the axial current.
The  Adler-Bardeen\cite{AB69}'s theorem says that higher-order effects in the triangular diagram of pion decay into two photons can be incorporated in the renormalization 
\[
\partial_\mu j_{\mu}^5=\epsilon^{\alpha\beta\mu\nu} F_{\alpha\beta}F_{\mu\nu} \frac{e_0^2}{8\pi^2}(1-\frac{3e_0^4}{64\pi^4}\log\frac{\Lambda^2}{m^2})
\]

The axial anomaly is not exhausted by the single loop, and radiative corrections were evaluated by several authors\cite{AI89, Ioffe06, Ioffe08}. Ioffe calculated the divergence of the axial current from the triangle diagram using the gluonic field strength as
\[
\partial_\mu j_\mu^{5\lambda}=-\frac{\alpha_s}{4\pi}\epsilon^{\alpha\beta\mu\nu}G_{\alpha\beta}G_{\mu\nu}\cdot N_c
\] 
where $N_c=3$ is the number of colors.

 The decay process of Nambu-Goldstone boson $\pi^0$ can be derived from current algebra and constrained by the symmetry of the vector current, whose triality sector is fixed. The gluons exchanged between the quark triangle, and the quark rectangle would be constrained to be in a triality sector, and the twisted diagram does not contribute. 
The standard chiral effective theory predicts\cite{Ioffe06}
\[
\Gamma(\pi^0\to \gamma\gamma)=\frac{\alpha^2}{32\pi^3}\frac{{m_\pi}^3}{{f_\pi}^2}=7.7{\rm eV}.
\]
Experimentally, $\Gamma(\pi^0\to \gamma\gamma)=7.82\pm 0.14\pm 0.17$eV\cite{PDG12}. 

%%%%%%
Divergence of an isoscalar axial current could play a role in $\eta\to \gamma\gamma$ and/or $\eta'\to \gamma\gamma$ decay processes, in which instanton contribute and the twisted diagram appears.  
A simple theoretical extension of $\pi^0\to \gamma\gamma$ to $\eta\to \gamma\gamma$ predicts\cite{Ioffe06}
\[
\Gamma(\eta\to \gamma\gamma)=\frac{\alpha^2}{32\pi^3}\frac{1}{3}\frac{{m_\eta}^3}{{f_\eta}^2}=0.13{\rm keV}
\]
Experimentally, $\Gamma(\eta\to \gamma\gamma)=0.510\pm 0.026$keV\cite{PDG12} is about 4 times larger.

A simple multiplication of $(m_{\eta'}/m_{\eta})^3$ predicts
\[
\Gamma(\eta'\to \gamma\gamma)=\frac{\alpha^2}{32\pi^3}\frac{1}{3}\frac{{m_{\eta'}}^3}{{f_{\eta'}}^2}=0.69{\rm keV}
\]
Experimental data of $\Gamma(\eta'\to \gamma\gamma)=4.34\pm 0.14$keV\cite{PDG12} is about six times larger, and the average decay width of $\eta$ and $\eta'$ is 5 times larger.

In \cite{DHL85},  one loop correction to the Wess-Zumino Lagrangean governing $\pi^0,\eta\to \gamma\gamma$ decay widths were studied. They used the eighth member of SU(3) octet, $\eta_8$ and the SU(3) singlet $\eta_0$ as
\begin{eqnarray}
\left|\eta\right\rangle&=&\cos\theta \left|\eta_8\right\rangle-\sin\theta\left|\eta_0\right\rangle\nonumber\\
\left|\eta'\right\rangle&=&\sin\theta \left|\eta_8\right\rangle+\cos\theta \left|\eta_0
\right\rangle\end{eqnarray}
and obtained $\theta\simeq-20^\circ$.
Ref. \cite{KL00,BN03} also obtained $\theta=-20^\circ$, but the studies of vacuum using quark-flavor basis rather than the octet-singlet basis turned out to fit data of $\eta,\eta'$ better\cite{Shore01, EF05}. 

The one-mixing angle scheme turned out to be valid only in large $N_c$ limit\cite{KL00,BB01,BN03}.
Bhagwat et al\cite{BCLRT07} defined the kernel of Bethe-Salpeter equation of pseudoscalar mesons as
\[
K=K_L+K_A
\] 
where $K_L$ is the reading order term and $K_A$ is the term associated with the anomaly, which is parametrized as
\begin{eqnarray}
&&(K_A){^{tu}_{rs}}(q,p,P)=-\xi((q-p)^2)\left\{\cos^2\theta_\xi[\zeta\, \gamma_5]_{rs}[\zeta\, \gamma_5]_{tu}
+\sin^2\theta_\xi[\zeta\,\gamma\cdot P\gamma_5]_{rs}[\zeta\,\gamma\cdot P\,\gamma_5]_{tu} \right\}\nonumber\\
&&\xi(k^2)=(2\pi)^4\xi\, \delta^4(k)
\end{eqnarray}
with a dimensionless coupling constant $\xi$ as a parameter, and the quark mass function
\begin{equation}
\zeta=diag[\frac{1}{M_u^D}, \frac{1}{M_d^D}, \frac{1}{M_s^D},\cdots], \quad M_f^D=M_f(s=0)
\end{equation}
where $f$ is the quark flavor. 

When the basis of non-strange $\eta$ and strange $\eta$,
\begin{eqnarray}
|\eta_{NS}\rangle&=&\frac{1}{\sqrt 2}(|u\bar u\rangle+|d\bar d\rangle)=\frac{1}{\sqrt 3}\left|\eta_8\right\rangle+\sqrt{\frac{2}{3}}\left|\eta_0 \right\rangle
\nonumber\\
|\eta_S\rangle&=&|s\bar s\rangle=-\sqrt{\frac{2}{3}}\left|\eta_8\right\rangle+\frac{1}{\sqrt 3}\left|\eta_0\right\rangle
\end{eqnarray}
are used,
\begin{eqnarray}
\left|{\eta}\right\rangle&=&\cos\phi \left|\eta_{NS}\right\rangle-\sin\phi\left|\eta_S\right\rangle\nonumber\\
\left|{\eta'}\right\rangle&=&\sin\phi \left|\eta_{NS}\right\rangle+\cos\phi\left|\eta_S\right\rangle.\label{eta_S}
\end{eqnarray}
The experimental value $\phi=41.88^\circ$ corresponds to $\left|\eta_{NS}\right\rangle=0.427, \left|\eta_S\right\rangle=0.0860$, and  $\theta=\phi-arctan \sqrt 2=\phi-54.74^\circ=-12.9^\circ$\cite{KK06}. 
When the parameter $\xi=0.076$ was taken, $\theta_\eta=-15.4^\circ$ and $\theta_\eta'=-15.7^\circ$ were obtained\cite{BCLRT07}. 

 In a phenomenological model using
the quark-flavor basis and assuming  mixing angle $\phi_q\simeq\phi_s\sim 40^\circ$, the decay width of  $\eta$ and $\eta'$ could be fitted\cite{EF05}, and a lattice simulation using the similar bases of (\ref{eta_S}), and including the loop of $c$ quarks, obtained also $\phi=46^\circ$\cite{MOU13}.

\newpage
\section{Triality symmetry of \'E.Cartan}
  \'E.Cartan\cite{Cartan66} studied algebra of system of spinors and vectors, which have the triality symmetry. 
He considered in the euclidean space $E_{2\nu}$, the semispinors $\phi$ which are specified by an even number of indices: $\xi_0,\xi_{23},\xi_{31},\xi_{12},\xi_{1234},\xi_{14},\xi_{24}$ and $\xi_{34}$
and semispinors $\psi$ which are specified by an odd number of indices:

The spinor bases $A,B,C,D$ are expressed as
\begin{eqnarray}
 &&A= \xi_{14}\sigma_x+\xi_{24}\sigma_y+\xi_{34}\sigma_z+\xi_{0} {\bf I}\nonumber\\
 &&B= \xi_{23}\sigma_x+\xi_{31}\sigma_y+\xi_{12}\sigma_z+\xi_{1234} {\bf I}\nonumber\\
 &&C= \xi_{1}\sigma_x+\xi_{2}\sigma_y+\xi_{3}\sigma_z+\xi_{4} {\bf I}\nonumber\\
 &&D= \xi_{234}\sigma_x+\xi_{314}\sigma_y+\xi_{124}\sigma_z+\xi_{123} {\bf I}
\end{eqnarray}
and the vector fields are expressed as
\begin{eqnarray}
&&E=x_1 {\Vec i}+x_2{\Vec j}+x_3{\Vec k}+x_4{\bf I}\nonumber\\
&&E'=x_1' {\Vec i}+x_2'{\Vec j}+x_3'{\Vec k}+x_4'{\bf I}.
\end{eqnarray}
The coupling of spinors to vector particles $x_i$ and $x_j'$ are expressed as ${^t\phi}CX\psi$. \'E. Cartan considered 5 superspace transformations $G_{23},G_{12},G_{13},G_{123}$ and $G_{132}$. By the transformation $G_{12}$, the spinor ${^t(}A,B)$ is transformed to $E,E'$ in which the 4th component of the vector field $x_4$ and $x_4'$ are interchanged, and $E,E'$ are transformed to ${^t(}A,B)$ in which $\xi_{1234}$ and $\xi_0$ are interchanged. 

The operator $G_{23}$ does not transform vectors to spinors or spinors to vectors,
but its operation on Dirac fermions is a charge conjugation,  
\begin{eqnarray}
&&G_{23}\left(\begin{array}{c} A\\
                           B\end{array}\right)=\left(\begin{array}{c}C\\
                                                                               D\end{array}\right), \qquad
G_{23}\left(\begin{array}{c} C\\
                                      D\end{array}\right)=-\left(\begin{array}{c}A\\
                                                                                           B\end{array}\right).\nonumber
\end{eqnarray}
and its operation on vector field $E$ and $E'$ is an interchange of the 4th component.
\begin{eqnarray}
&&G_{23}\left(\begin{array}{c} E\\
                           E'\end{array}\right)=G_{23}\left(\begin{array}{c}x_1{\bf i}+x_2{\bf j}+x_3{\bf k}+x_4{\bf I}\\
                                        x'_1{\bf i}+x'_2{\bf j}+x'_3{\bf k}+x'_4{\bf I}\end{array}\right)\nonumber\\
&&=\left(\begin{array}{c}x_1{\bf i}+x_2{\bf j}+x_3{\bf k}-x'_4{\bf I}\\
                                        x'_1{\bf i}+x'_2{\bf j}+x'_3{\bf k}-x_4{\bf I}\end{array}\right).\nonumber
\end{eqnarray}
When the vector fields are selfdual, and the energy is null, a singular behavior can be expected from $G_{23}E$ 
and $G_{23}E'$.   

When $G_{13}$ operates on left-handed fermion the 4th component of $A$ and $B$, $\xi_0$ and $\xi_{1234}$, respectively are interchanged:

\begin{eqnarray}
&&G_{13}\left(\begin{array}{c} A\\
                           B\end{array}\right)=G_{13}\left(\begin{array}{c}\xi_{14}\sigma_x+\xi_{24}\sigma_y+\xi_{34}\sigma_z+\xi_0{\bf I}\\
                              \xi_{23}\sigma_x+\xi_{31}\sigma_y+\xi_{12}\sigma_z+\xi_{1234}{\bf I}\end{array}\right),\nonumber\\
&&=
\left(\begin{array}{c} \xi_{14}\sigma_x+\xi_{24}\sigma_y+\xi_{34}\sigma_z+\xi_{1234}{\bf I}\\
                              \xi_{23}\sigma_x+\xi_{31}\sigma_y+\xi_{12}\sigma_z+\xi_0{\bf I}\end{array}\right)
\nonumber.
\end{eqnarray}

\[
G_{13}\left(\begin{array}{c}E\\
                               E'\end{array}\right)=
\left(\begin{array}{c} D\\
                           C\end{array}\right)\nonumber
=\left(\begin{array}{c}\xi_{234}\sigma_x+\xi_{314}\sigma_y+\xi_{124}\sigma_z+\xi_{123}{\bf I}\\
                                         \xi_{1}\sigma_x+\xi_{2}\sigma_y+\xi_{3}\sigma_z+\xi_4{\bf I}\end{array}\right)
\]

In quantum electrodynamics, the success of Dirac spinors which are described by quaternions is established, but in QCD, it is not so evident.
I study in this paper, consequences of the interchange of the 4th component in the octonion by
extending the QCD using octonions.
The QCD in quaternion basis was studied in \cite{SF09,SF10,SF11,SF12a,SF12b}, and recently,
I discussed the axial anomaly using octonion bases and considered rescattering diagrams in \cite{SF13}.

The vector particles that propagate between the triangle diagram and the square diagram can be photons as well as gluons. In low energy, the effect of gluons is expected to be important, but transition from two gluons to two photons via square diagram or the gluon and photon scattering at low energy is not well understood\cite{BINT91}.  

In the rescattering diagram, I considered the change of vector particles $x_i$ and $x_j$ to $x_i'$ and $x_j'$ without changing the polarization direction. 

In chiral gauge theory, left-handed quarks and left-handed leptons are defined as
\[
Q_L=\left(\begin{array}{c}u\\
                                 d\end{array}\right)\qquad L_L=\left(\begin{array}{c}\nu\\
                                                                                                   l\end{array}\right)
\]
and right-handed leptons $\Psi_R$ are defined as
\[
\Psi_{L_i}'=\sigma^2 \Psi_{R_i}^* \qquad \Psi_{L_i}'^\dagger=\Psi_{R_i}^T\sigma^2
\]

The left-handed fermion ${^t(}A, B)$ is defined as a fermion and its right-handed anti-particle $(C,D)$ is defined as a new right-handed fermion.

The fermionic Lagrangian in the standard model is
\[
{\mathcal L}=\bar\Psi i\gamma^\mu \left(\partial_\mu-ig A^a_\mu t^a_r\left (\frac{1-\gamma^5}{2}\right)\right)
\Psi
\]
In \'E. Cartan's convention, the projection operator $\frac{1-\gamma^5}{2}$ is contained in $\phi$ or $\psi$ and
\[
\bar\Psi=(\phi,\psi)\left(\begin{array}{cc}0&1\\
                                                      1&0\end{array}\right) =(\psi,\phi)                                                                                 
\]

Quarks and leptons belong to their triality sectors, and I assume that the electromagnetic interaction has the triality selection rules, or  electron $e$, muon $\mu$ and tauon $\tau$  react to quarks in their same triality sector. Quarks may emit photons in an arbitrary triality sector, but if photons from quarks in the triality sector different from that of leptons are not detected,  and if $\mu\to e\gamma$ is not observed, ( its actual  experimental probability is less than $10^{-11}$), the triality selection rule can be established. 

The vector field $E$ and $E'$ of \'E. Cartan is a kind of generalization of the electro magnetic field $A_\mu$.
When a pair of vector particles $E,E$ or $E'E'$ transform into a quark-antiquark pair and return to two same type of vectors $E,E$ or $E',E'$, we call the diagram a rescattering diagram. The quark-antiquark system may
have isospin 1, like $\pi$. 
When different types $E,E'$ transform into a quark-antiquark pair and return to $E, E'$, we call the diagram  a twisted diagram. The corresponding quark-antiquark systems have isospin 0, like $\eta$ and $\eta'$.

\section{Rescattering diagrams}

 Due to the $\gamma_5$ operator in the axial vector vertex, the triangle diagram introduces an operator containing  4 components  $\partial_i,x_j,\partial_k$ and $x_m$, combined by the anti-commutation factor $\epsilon_{ijkm}$.  Emitted two vector particles can be absorbed by a fermion, and the fermion can emit two vector particles. When the polarizations of the two vector particles do not change in the two emissions, I call the diagram as a rescattering diagram.

In QED, a $e_0^6$ term appears from rescattering of two photons of two different polarizations, as shown in Figs.1-8. They contain square diagrams contoured by four quark lines. In Figs.1-4, axial vector current couples to fermions with an even number of indices, and In Figs.5-8, it couples to fermions with an odd number of indices.

I choose the propagator between two vector particles emission points in the $A_1 x_2 x_3$ or $A_1 x_2' x_3'$ diagram and between the final $x_2' x_3'$ or $x_2 x_3$ emission points to be spinless.   In this model,  the $A_1(A_1')\sigma_x$ vertex is produced from the product:

 $\xi_{12}\sigma_z\times \xi_{31}\sigma_y\to -A_1\sigma_x$, $\xi_{34}\sigma_z\times \xi_{24}\sigma_y\to -A_1\sigma_x$ ,
$\xi_3\sigma_z\times \xi_2\sigma_y\to -A_1'\sigma_x$\\ and $\xi_{124}\sigma_z\times \xi_{314}\sigma_y\to -A_1'\sigma_x$ .

Similarly,  $A_2(A_2')\sigma_y$ vertex is produced from the product:

 $\xi_{23}\sigma_x\times \xi_{12}\sigma_z\to -A_2'\sigma_y$, $\xi_{14}\sigma_x\times \xi_{34}\sigma_z\to -A_2'\sigma_y$ ,
$\xi_1\sigma_x\times \xi_3\sigma_z\to -A_2\sigma_y$\\ and  $\xi_{124}\sigma_z\times \xi_{234}\sigma_x\to A_2\sigma_y$ 

Similarly,  $A_3(A_3')\sigma_z$ vertex is produced from the product:

$\xi_{31}\sigma_y\times \xi_{23}\sigma_x\to -A_3'\sigma_z$, $\xi_{24}\sigma_y\times \xi_{14}\sigma_x\to -A_3'\sigma_z$ ,
$\xi_2\sigma_y\times \xi_1\sigma_x\to -A_3\sigma_z$\\ and  $\xi_{314}\sigma_y\times \xi_{234}\sigma_x\to -A_3\sigma_z$ 

\begin{figure}[htb]
\begin{minipage}[htb]{0.47\linewidth}
\begin{center}
\includegraphics[width=6cm,angle=0,clip]{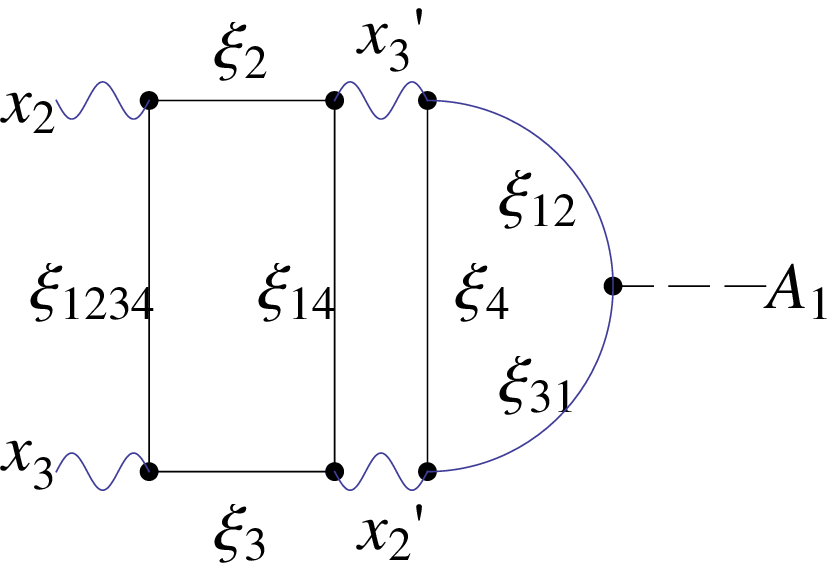}
\caption{The half circle diagram of an axial anomaly. $A_1x_3' x_2'$ type and its rescattering.} 
\label{g1d}
\end{center}
\end{minipage}
\hfill
\begin{minipage}[htb]{0.47\linewidth}
\begin{center}
\includegraphics[width=6cm,angle=0,clip]{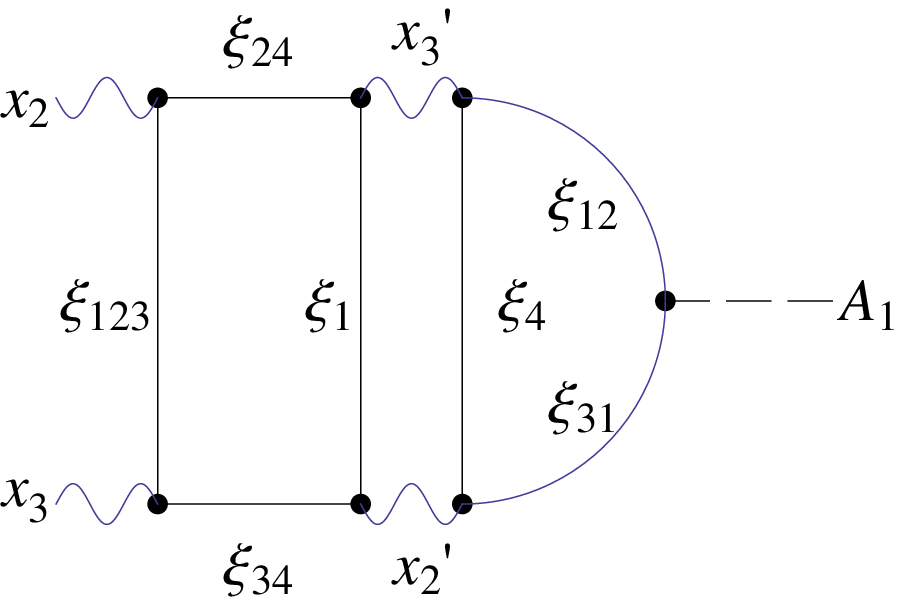}
\caption{The half circle diagram of an axial anomaly. $A_1 x_3' x_2'$ type and its rescattering.}
\label{g1f}
\end{center}
\end{minipage}
\end{figure}
\begin{figure}
\begin{minipage}[htb]{0.47\linewidth}
\begin{center}
\includegraphics[width=6cm,angle=0,clip]{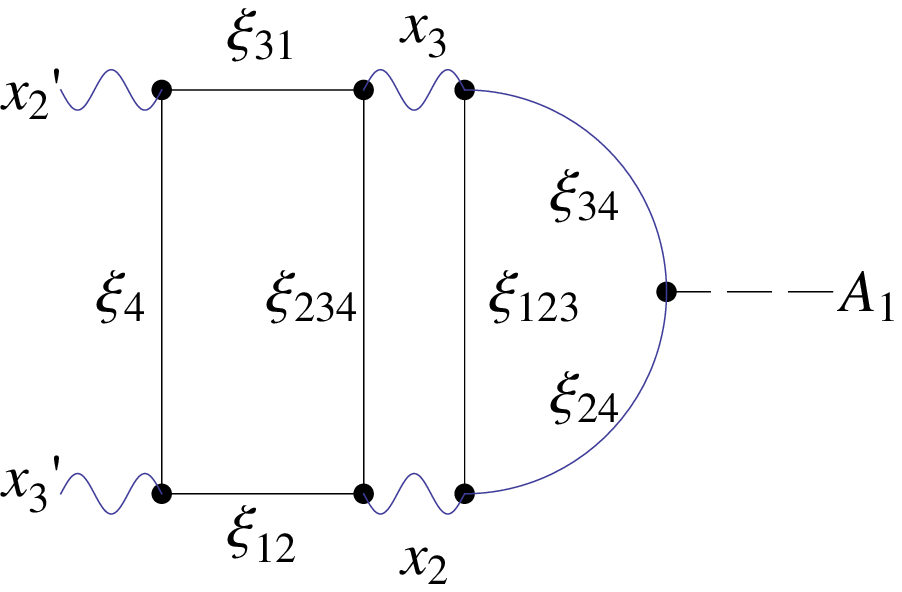}
\caption{%The half circle diagram of an axial anomaly.
$A_1 x_3 x_2$ type and its rescattering.} 
\label{g1h}
\end{center}
\end{minipage}
\hfill
\begin{minipage}[htb]{0.47\linewidth}
\begin{center}
\includegraphics[width=6cm,angle=0,clip]{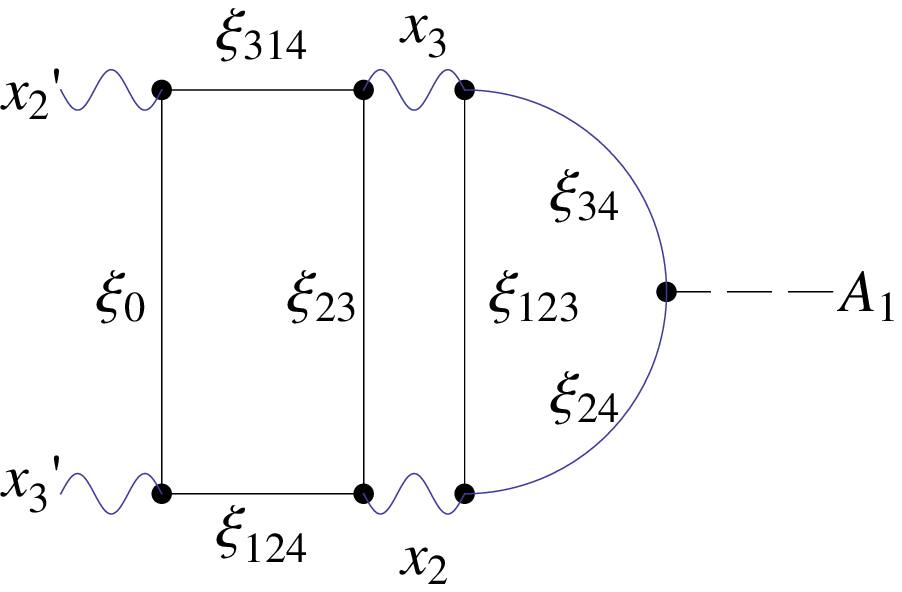}
\caption{%The half circle diagram of an axial anomaly. 
$A_1 x_2 x_3$ type and its rescattering.}
\label{g1i}
\end{center}
\end{minipage}
\end{figure}
\begin{figure}
\begin{minipage}[htb]{0.47\linewidth}
\begin{center}
\includegraphics[width=6cm,angle=0,clip]{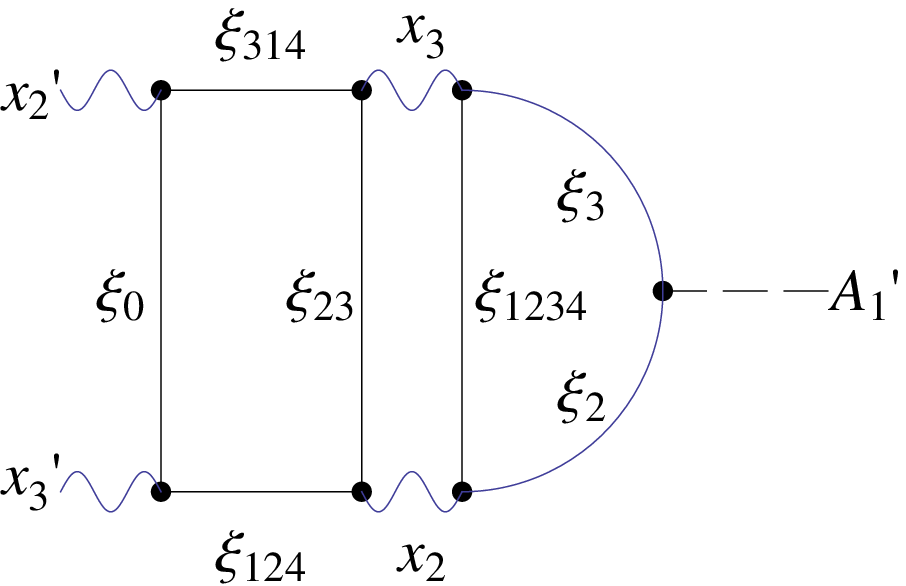}
\caption{%The half circle diagram of an axial anomaly. 
$A_1' x_3 x_2$ type and its rescattering.} 
\label{g1ap}
\end{center}
\end{minipage}
\hfill
\begin{minipage}[htb]{0.47\linewidth}
\begin{center}
\includegraphics[width=6cm,angle=0,clip]{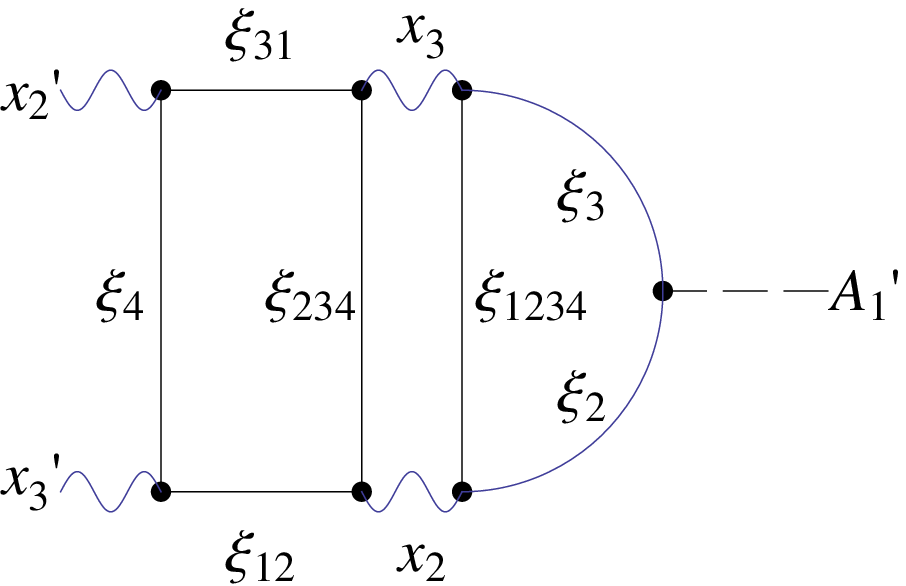}
\caption{%The half circle diagram of an axial anomaly. 
$A_1' x_2 x_3$ type and its rescattering.}
\label{g1ep}
\end{center}
\end{minipage}
\end{figure}
\begin{figure}
\begin{minipage}[htb]{0.47\linewidth}
\begin{center}
\includegraphics[width=6cm,angle=0,clip]{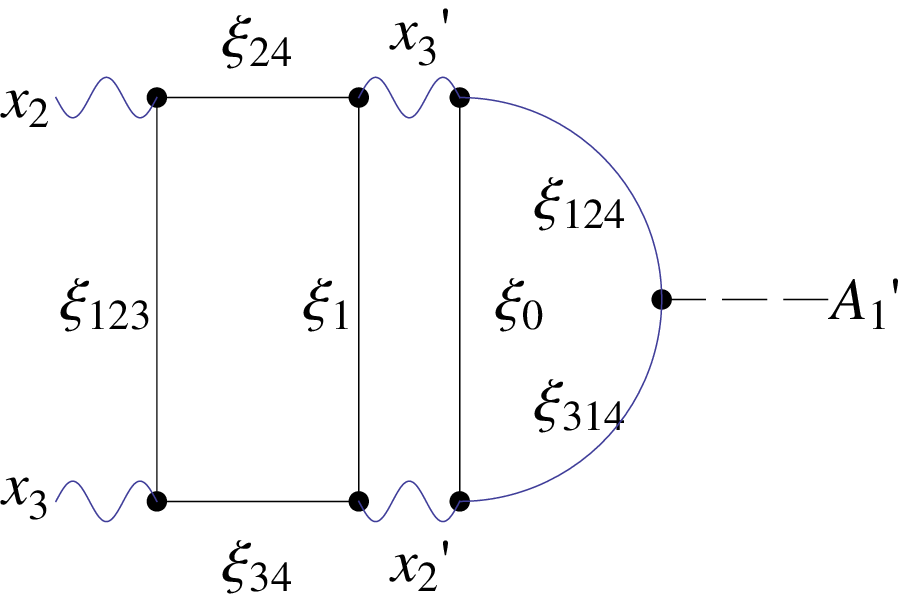}
\caption{%The half circle diagram of an axial anomaly. 
$A_1' x_3' x_2'$ type and its rescattering.} 
\label{g1b}
\end{center}
\end{minipage}
\hfill
\begin{minipage}[htb]{0.47\linewidth}
\begin{center}
\includegraphics[width=6cm,angle=0,clip]{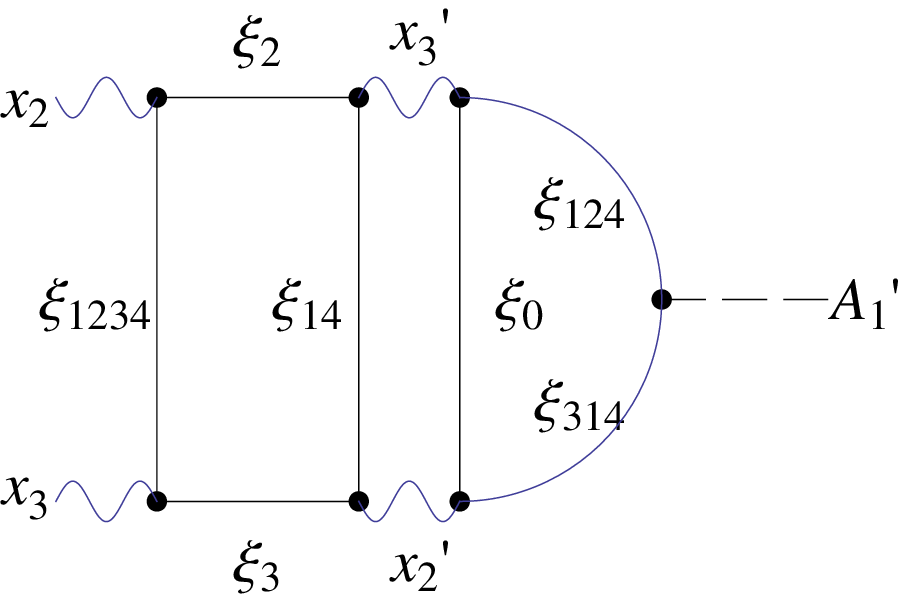}
\caption{%The half circle diagram of an axial anomaly. 
$A_1' x_2' x_3'$ type and its rescattering.}
\label{g1g}
\end{center}
\end{minipage}
\end{figure}

The amplitude that includes the half circle of Fig.1 and Fig.2 is
\begin{eqnarray}
T_{321}'(k,q)&=&-e^2\int\frac{d^4p}{(2\pi)^4}Tr [ \gamma_3 \frac{1}{p-(k-q)-m}\gamma_1\gamma_5\frac{1}{p-q-m}\gamma_2\frac{1}{p-k-m}]
\nonumber\\
&\propto& G'_{3\sigma}(k)G'_{2\rho}(q)\epsilon^{3\sigma 2\rho}
\end{eqnarray}
where $G'_{\lambda\mu}(k)=k_\lambda x'_\mu-k_\mu x'_\lambda$, and $\epsilon^{3124}=1$ and $\epsilon^{3421}=-1.$

The amplitude that includes the half circle of Fig.3 and Fig.4 is
\[
T_{321}(k,q)\propto G_{3\sigma}(k)G_{2\rho}(q)\epsilon^{3\sigma 2\rho}
\]
where $G_{\lambda\mu}(k)=k_\lambda x_\mu-k_\mu x_\lambda$.

%\newpage
\section{Twisted diagrams}

  When the vector particles are self-dual, there is no reason to restrict an emission of two vector particles in the triangle diagram to be of type $x_i x_j$ or $x_i' x_j'$. 
When the emitted  vector particles are $x_4 x_i'$ or $x_4' x_i$, where $i=1,2,3$, the fermion square diagram can emit two vector particles $x_j x_k'$ or $x_j' x_k$ where $j\ne k$ and they are not equal to $ 'i'$ and '4'.  When the emitted vector particles are $x_2' x_3$ or $x_2 x_3'$, the fermion square can emit $x_1 x_4'$ or $x_1' x_4$, respectively.

When the vector fields can be treated as self-dual, i.e. $x_i$ and $x_j'$ interact with $\xi_{1234}$ and $\xi_0$, one could consider topologically complicated processes in which after twice rotations, (the first via $\xi_{1234}$ and the second via $\xi_0$)  the initial configuration reappears.

$G_{13}$ transforms $(A_1,A_2,A_3,A_4)\to(A_1,A_2,A_3,B_4)$ and $(B_1,B_2,B_3,B_4)\to (B_1,B_2,B_3,A_4)$. In other words, $\xi_{1234}$ is replaced by $\xi_0$. Since the intermediate quark propagator that absorbs and emits $x_1 x_4'$ or $x_4 x_1'$ is not expected to be in the eigenstate of $G_{13}$, the quark $\xi_{123}$ that absorbs $x_4'$ is transformed to $\xi_0$ but through the mixed component of $\xi_{1234}$,  it will emit $x_1$ and changes to $\xi_1$ as shown in Fig.9.
 
When vector particles are selfdual and $x_i$ and $x_i'$ are indistinguishable, the square in the left hand side of Fig.9 and Fig.10 give an amplitude, in which a spinor $\xi_a$ emits a vector particle $x_b$ and changes to a spinor $\xi_c$ is represented as $\xi_a, x_b, \xi_c$, of the type
\[
(\xi_{1234},x_4'), \xi_1, x_3', \xi_{24}, x_2, \xi_{123}, x_4', \xi_0, x_1,
\]
\[
(\xi_0, x_1), \xi_{123}, x_3,\xi_{34}, x_2', \xi_1, x_1, \xi_{1234}, x_4'.
\]
They emit $x_2, x_2'$ and $x_3, x_3'$ but do not contain  $x_4$ and $x_1'$.  

 They are included in the amplitudes
\[
T_{3241}(K,Q,q,k)\propto G_{3\rho}'(K) G_{2\tau}(Q) G_{4\upsilon}(q)' G_{1\sigma}(k) \epsilon^{ 3\rho2\tau}\epsilon^{ 4\upsilon 1\sigma}
\]
\[
T_{3214}(K,Q,k,q)\propto G_{3\rho}(K) G_{2\tau}'(Q) G_{1\sigma}(k)G_{4\upsilon}'(q) \epsilon^{ 3\rho2\tau }\epsilon^{1\sigma 4\upsilon}.
\]

The amplitude that includes the right hand loop of Fig.9 and Fig.10 are
\begin{eqnarray}
T_{41}''(k,q)&\propto& G_{4\overline\upsilon}'(q) G_{1\overline\sigma}(k)\epsilon^{ 4\overline\upsilon 1\overline\sigma}
\nonumber\\
T_{14}'''(k,q)&\propto& G_{1\overline\sigma}' (k)G_{4\overline\upsilon}(q)\epsilon^{1\overline\sigma 4\overline\upsilon}.
\end{eqnarray}

 They are included in the amplitudes
\begin{eqnarray}
T_{324141\mu}(K,Q,q,k))&=&-e^2g^4 \int\frac{d^4p}{(2\pi)^4}  \int\frac{d^4P}{(2\pi)^4}Tr [ \gamma_3 \frac{1}{(P+K)-m}\gamma_2\frac{1}{(P+K+Q)-m}\nonumber\\
&& \times\gamma_1\frac{1}{P-m} \gamma_4\frac{1}{(P+k)-m}]\frac{1}{k^2+\epsilon}\frac{1}{q^2+\epsilon}
\delta(P+K+Q-p-k-q)\nonumber\\
&& \times Tr[\gamma_4 \frac{1}{(p-k+q)-m}\gamma_\mu\gamma_5\frac{1}{p-m} \gamma_1\frac{1}{(p-k)-m}] \nonumber\\
&\propto& G_{3\rho}'(K) G_{2\tau}(Q) G_{4\upsilon}'(q) G_{1\sigma}(k) \epsilon^{ 3\rho 2\tau}\epsilon^{ 4\upsilon 1\sigma} G_{4\overline\upsilon}'(q)  G_{1 \overline\sigma}(k) \epsilon^{ 4\overline\upsilon 1\overline\sigma}\nonumber\\
&\propto&G_{3\rho}'(K) G_{2\tau}(Q)  \epsilon^{ 3\rho 2\tau}( G_{43}'(q) G_{12}(k)-G_{42}'(q) G_{13}(k) )^2
\nonumber
\end{eqnarray}
\begin{eqnarray}
T_{321414\mu}(K,Q,k,q)&=&-e^2g^4 \int\frac{d^4p}{(2\pi)^4}  \int\frac{d^4P}{(2\pi)^4}Tr [ \gamma_3 \frac{1}{(P+K)-m}\gamma_2\frac{1}{(P+K+Q)-m}\nonumber\\
&& \times\gamma_1\frac{1}{P-m} \gamma_4\frac{1}{(P+q)-m}]\frac{1}{k^2+\epsilon}\frac{1}{q^2+\epsilon}
\delta(P+K+Q-p-q-k)\nonumber\\
&&\times Tr[ \gamma_4 \frac{1}{(p-q+k)-m}\gamma_\mu\gamma_5\frac{1}{p-m}\gamma_1\frac{1}{(p-q)-m} ]\nonumber\\
&\propto& G_{3\rho}(K) G_{2\tau}'(Q) G_{1\sigma}(k)G_{4\upsilon}'(q)\epsilon^{ 3\rho 2\tau }\epsilon^{1\sigma 4\upsilon}
G_{1\overline\sigma}(k)G_{4\overline\upsilon}'(q) \epsilon^{1\overline\sigma 4\overline\upsilon}\nonumber\\
&\propto&. G_{3\rho}(K) G_{2\tau}'(Q) \epsilon^{ 3\rho 2\tau} ( G_{13}(k) G_{4 2}'(q)-G_{12}(k) G_{43}'(q) )^2\nonumber
\end{eqnarray}

In Figs.9-16,  the quark spinor $\xi_{1234}I$ absorbs vectors $x_i$, the spinor $\xi_0 I$ absorbs $x'_i$. Gluons polarized in the four directions $x_1',x_2',x_3,x_4$, $x_1,x_2,x_3', x_4'$ or $x_1',x_2,x_3',x_4$ ( and $x_i, x_i'$ interchanged) appear on the twisted diagrams.

 The vector $x_4$ and $x_1'$ have common coupling to spinor $\xi_{123}$ and $\xi_1$.
 In Fig.9, the product $\xi_{1}{\sigma_x} \times \xi_{24}\sigma_y \to x_3'{\bf k}$ and $\xi_{24}\sigma_y\times \xi_{123}I\to x_2{\bf j}$ the final two vectors  $x_3'{\bf k}\times x_2{\bf j}$ makes a vector product -$x_1{\bf i}$.
After emission of $x_2$, $\xi_{123}$ emits a $x_4'$ and becomes $\xi_0$, and as shown in Fig.10,  absorbs $x_4'$ and becomes $\xi_{123}$. 
It emits $x_3$ and $x_2'$ and becomes $\xi_1$. It emits $x_1$ and goes to $\xi_{1234}$. $\xi_{1234}$ absorbs $x_1$
as shown in Fig.9 and emits $x_3'$ and $x_2$ and returns to $\xi_{123}$.
In Fig.11 and Fig.12, the spinor $\xi_{123}$ is replaced by $\xi_1$.

The relative sign of $x_3'{\bf k}\times x_2{\bf j}$ from Fig.9 and $x_2'{\bf j}\times x_3{\bf k}$ from Fig.10 is cancelled by the relative sign of the product of 6 vertices $\xi_* x_j \xi_{*'}$, where $j$ runs from 1 to 4, and $\xi_*,  \xi_{*'}$ run $\xi_\alpha, \xi_{\alpha\beta}, \xi_{\alpha\beta\gamma}$ or $\xi_{1234}$ that appear on the Figures.

 The vector $x_4'$ and $x_1$ have the common coupling  to spinor $\xi_{234}$ and $\xi_4$. 
In Figs. 13-16, the same mechanism occurs as in Figs.9-12, in which $\xi_{123}$ is replaced by $\xi_{234}$ and $\xi_1$ is replaced by $\xi_4$.

\begin{figure}[htb]
\begin{minipage}[b]{0.47\linewidth}
\begin{center}
\includegraphics[width=6cm,angle=0,clip]{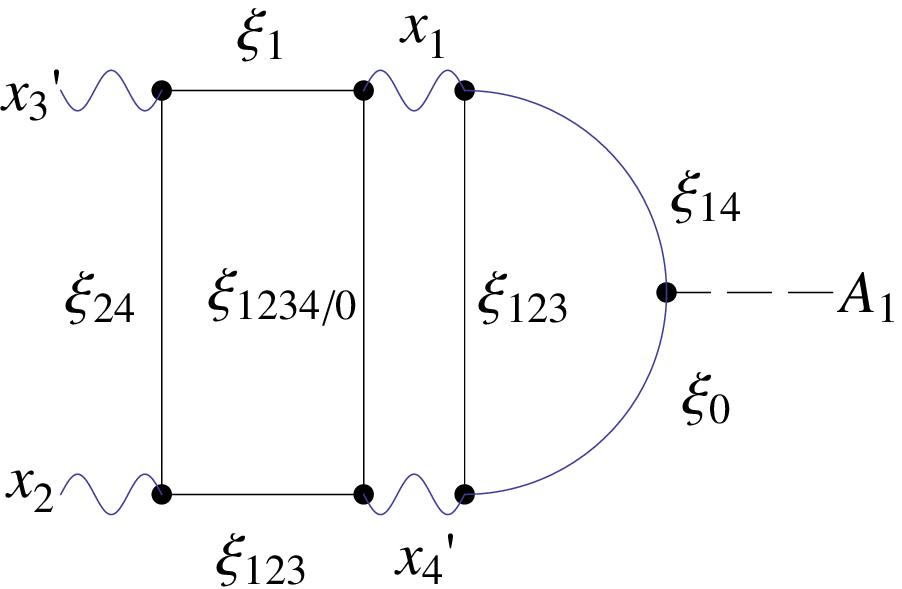}
\caption{The axial anomaly diagram via $A_1x_1 x_4'$ to $A_1 x_3' x_2$.} %9
\label{gb1e}
\end{center}
\end{minipage}
\hfill
\begin{minipage}[b]{0.47\linewidth}
\begin{center}
\includegraphics[width=6cm,angle=0,clip]{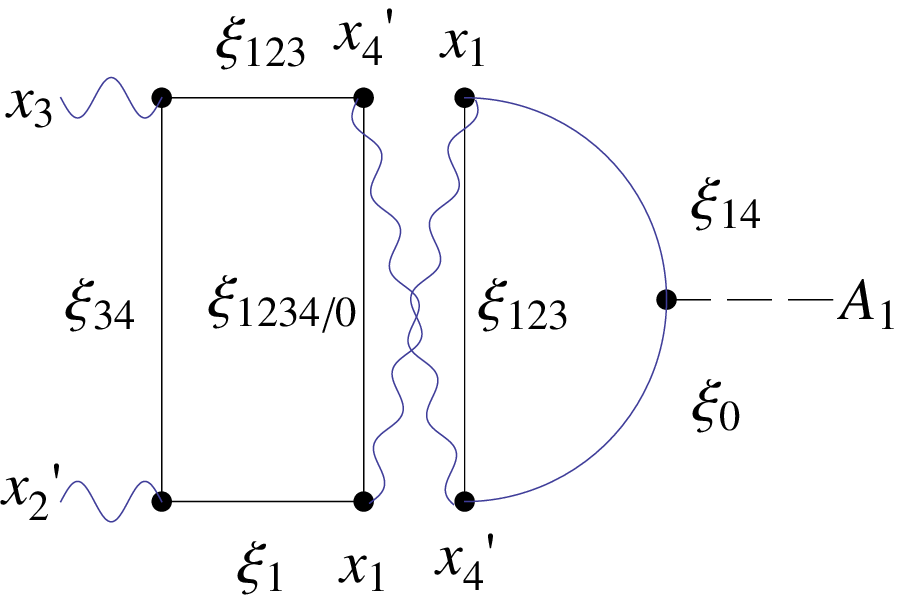}
\caption{The axial anomaly diagram via $A_1 x_1 x_4'$ to $A_1 x_3 x_2'$.}%10
\label{gb1ft}
\end{center}
\end{minipage}
\end{figure}

\begin{figure}[bottom]
\begin{minipage}[b]{0.47\linewidth}
\begin{center}
\includegraphics[width=6cm,angle=0,clip]{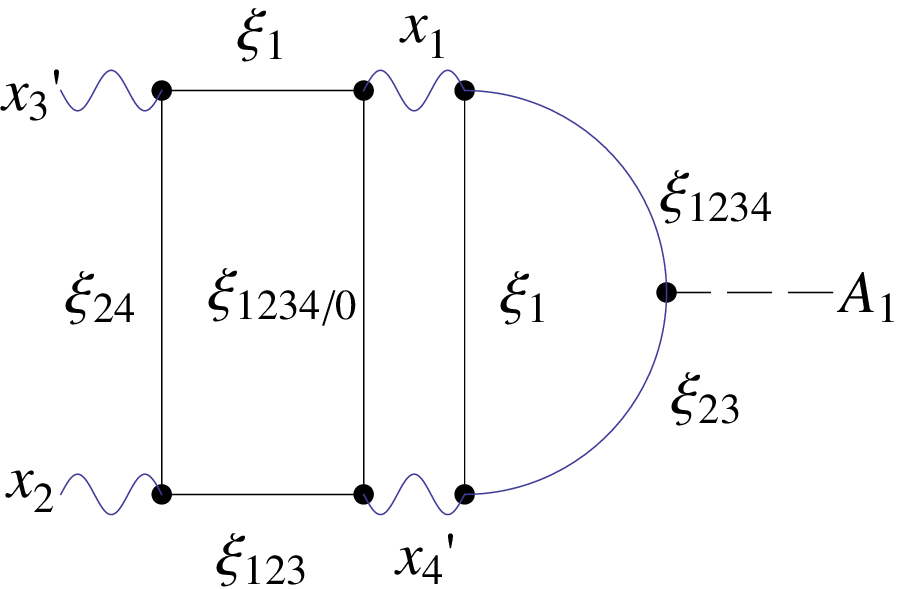}
\caption{The diagram via $A_1x_1 x_4'$ to $A_1 x_3' x_2$.} %9p
\label{gb1bp}
\end{center}
\end{minipage}
\hfill
\begin{minipage}[b]{0.47\linewidth}
\begin{center}
\includegraphics[width=6cm,angle=0,clip]{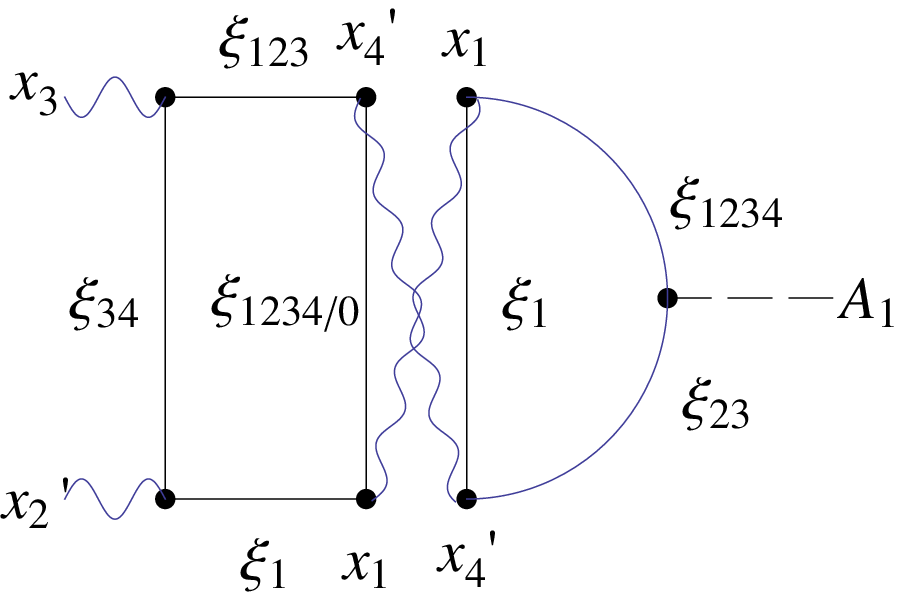}
\caption{The diagram via $A_1 x_1 x_4'$ to $A_1 x_3 x_2'$.}%10p
\label{gb1fpt}
\end{center}
\end{minipage}
\end{figure}

\begin{figure}
\begin{minipage}[b]{0.47\linewidth}
\begin{center}
\includegraphics[width=6cm,angle=0,clip]{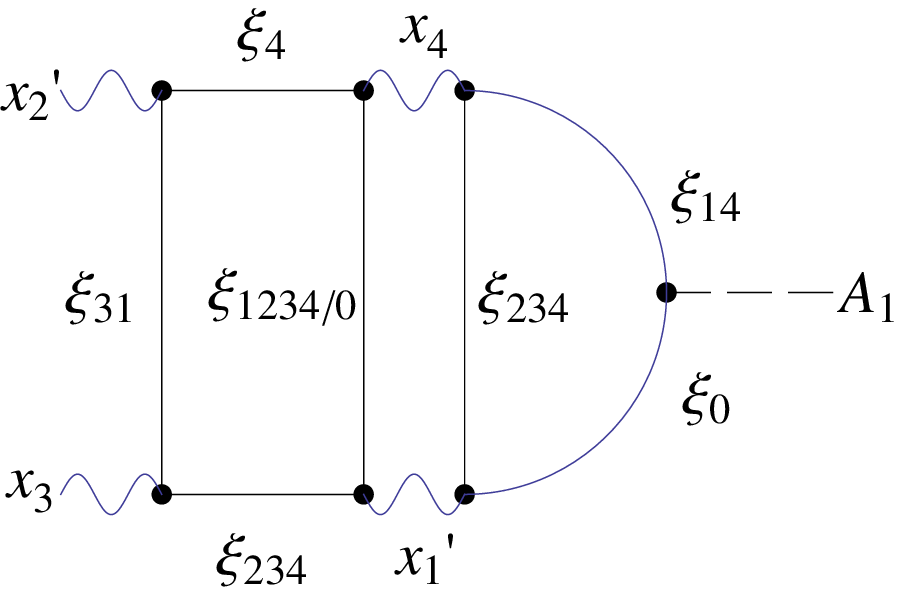}
\caption{The diagram via $A_1 x_4 x_1'$ to $A_1x_2' x_3$.} %11
\label{gb1a}
\end{center}
\end{minipage}
\hfill
\begin{minipage}[b]{0.47\linewidth}
\begin{center}
\includegraphics[width=6cm,angle=0,clip]{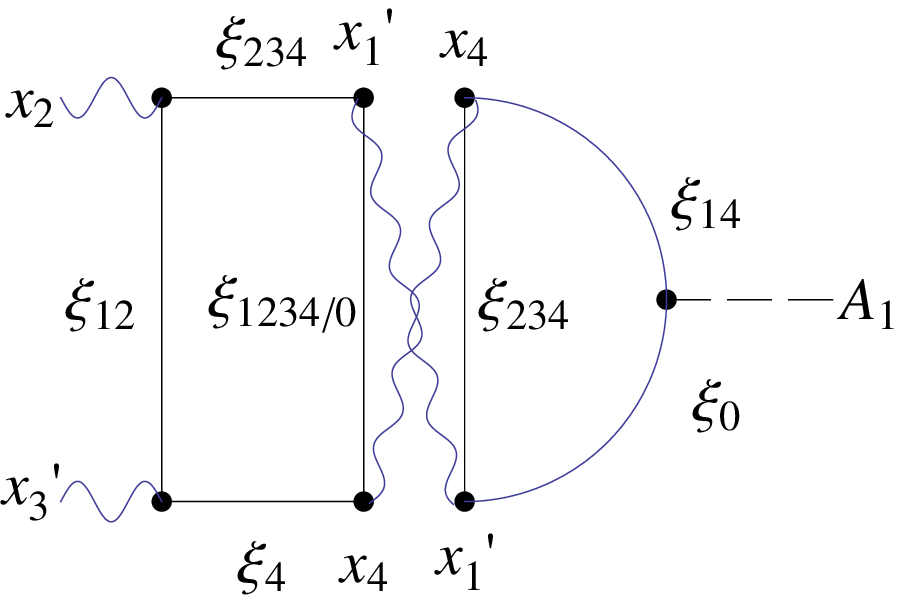}
\caption{\small The diagram via $A_1 x_4 x_1'$ to $A_1 x_2 x_3'$.}%12
\label{gb1dt}
\end{center}
\end{minipage}
\end{figure}

\begin{figure}
\begin{minipage}[b]{0.47\linewidth}
\begin{center}
\includegraphics[width=6cm,angle=0,clip]{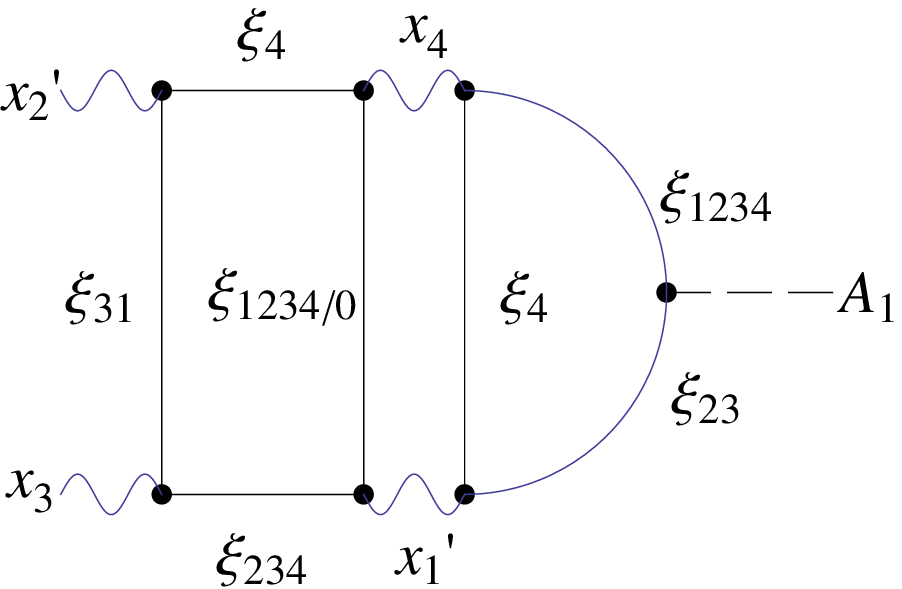}
\caption{The diagram via $A_1 x_4 x_1'$ to $A_1x_2' x_3$.} %11p
\label{gb1ap}
\end{center}
\end{minipage}
\hfill
\begin{minipage}[b]{0.47\linewidth}
\begin{center}
\includegraphics[width=6cm,angle=0,clip]{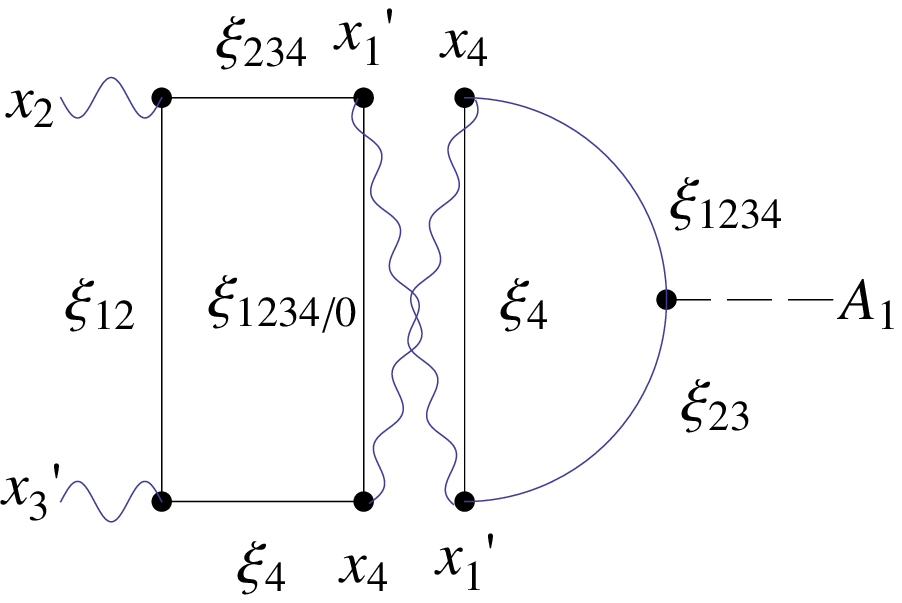}
\caption{\small The diagram via $A_1 x_4 x_1'$ to $A_1x_2 x_3' $.}%12p
\label{gb1dpt}
\end{center}
\end{minipage}
\end{figure}

Diagrams for vertices via $A_2 x_2 x_4'$ to $A_2 x_1' x_3$, etc. and via $A_3 x_3 x_4'$ to $A_3 x_1 x_2'$, etc. are similar.

In Figs.17-24, I show the cases of the vector particles  $x_1/x'_1$ and $x'_4/x_4$ in the final state.
In Figs.17-18, the quark propagator between $x_2'$ and $x_3$ is $\xi_{314}$ on the side of $A_1$ vertex, but $\xi_{1234}$ or $\xi_0$ on the side of $x_1 {x_4}'$. $\xi_{1234}$ and $\xi_0$ are the 4th component of spinor $B$ and $A$, respectively. When $G_{13}$ operates on ${^t(}A,B)$, $\xi_{1234}$ and $\xi_0$ are interchanged. Therefore, a mixing of $\xi_{1234}$ and $\xi_0$ inside the quark loop occur.

The Fig.17 and Fig.18 are contained in the amplitude
\begin{eqnarray}
T_{1432}(K,Q,k,q)&\propto& G_{1\rho}(K) G_{4\tau}'(Q) G_{3\upsilon}(k) G_{2\sigma}'(q) \epsilon^{1\rho 4\tau }\epsilon^{3\upsilon  2\sigma }\nonumber\\
T_{1423}(K,Q,q,k)&\propto& G_{1\rho}'(K) G_{4\tau}(Q) G_{2\sigma}'(q) G_{3\upsilon}(k)\epsilon^{ 1\rho 4\tau }\epsilon^{2\sigma 3\upsilon}.
\nonumber
\end{eqnarray}

Mixing of $\xi_{1234}$ and $\xi_0$ is important. Experimentally, ${x_1}' x_4$ or $x_1 {x_4}'$ state will not be detected, but via rescattering it changes to ${x_2}' x_3$ or $x_2 {x_3}'$ state.
\begin{figure}
\begin{minipage}[b]{0.47\linewidth}
\begin{center}
\includegraphics[width=6cm,angle=0,clip]{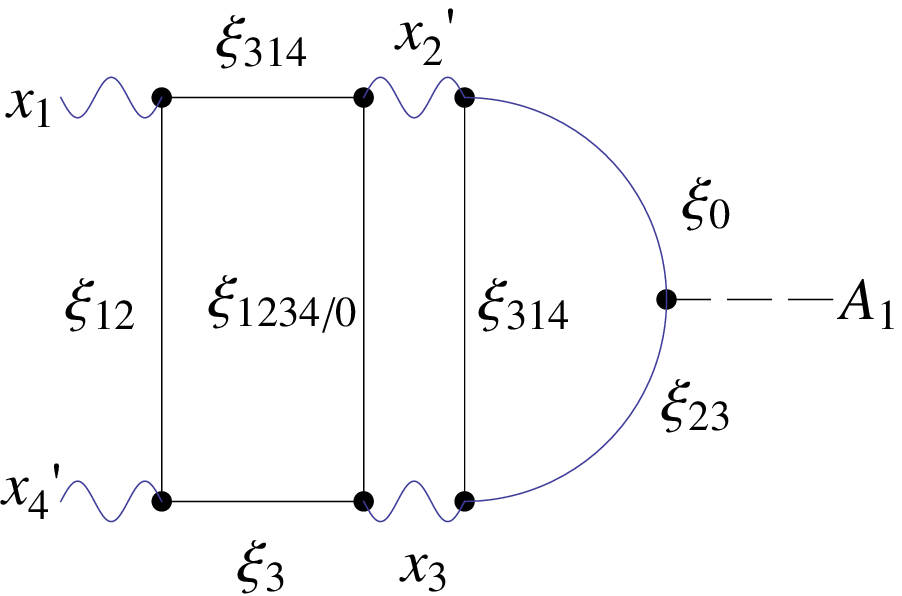}%21
\caption{The diagram via $A_1x_2' x_3$ to $A_1 x_1 x_4'$.}
\label{gb4a}
\end{center}
\end{minipage}
\hfill
\begin{minipage}[b]{0.47\linewidth}
\begin{center}
\includegraphics[width=6cm,angle=0,clip]{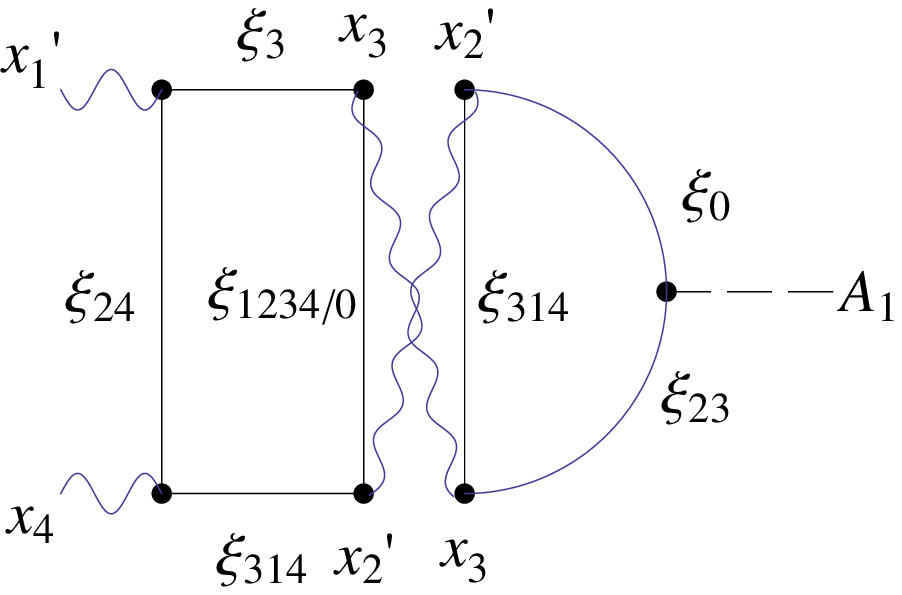}%22
\caption{The diagram via $A_1x_2' x_3$ to $A_1 x_1' x_4$.}
\label{gb4ct}
\end{center}
\end{minipage}
\end{figure}

\begin{figure}
\begin{minipage}[b]{0.47\linewidth}
\begin{center}
\includegraphics[width=6cm,angle=0,clip]{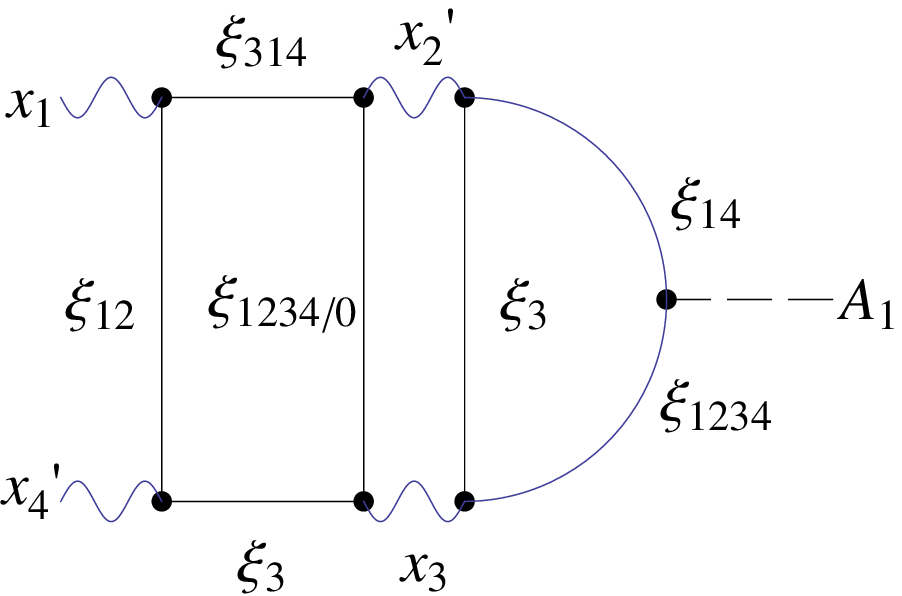}%21p
\caption{The diagram via $A_1 x_2' x_3$ to $A_1 x_1 x_4'$.} 
\label{gb4ap}
\end{center}
\end{minipage}
\hfill
\begin{minipage}[b]{0.47\linewidth}
\begin{center}
\includegraphics[width=6cm,angle=0,clip]{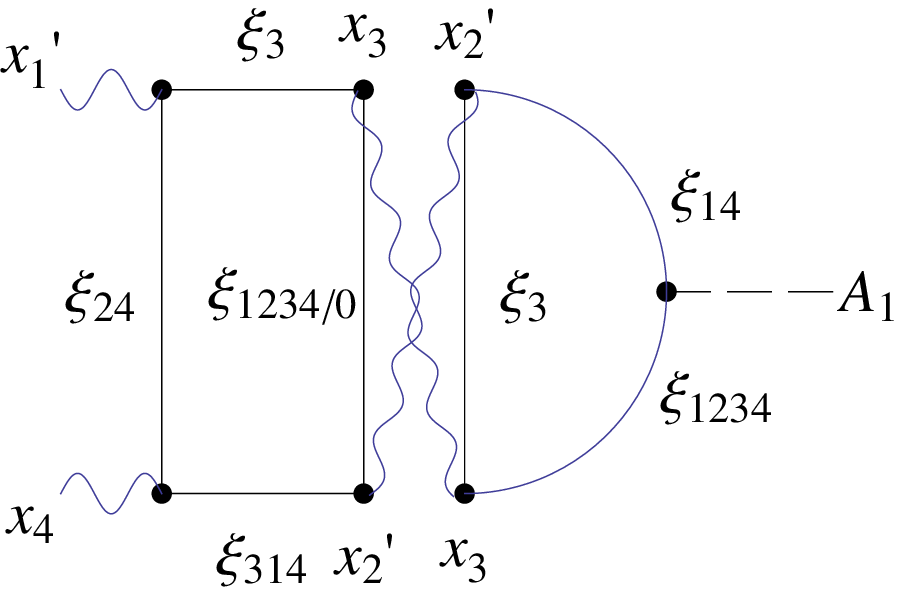}%22p
\caption{The diagram via $A_1 x_2' x_3$ to $A_1 x_1' x_4$.}
\label{gb4cpt}
\end{center}
\end{minipage}
\end{figure}

\begin{figure}
\begin{minipage}[b]{0.47\linewidth}
\begin{center}
\includegraphics[width=6cm,angle=0,clip]{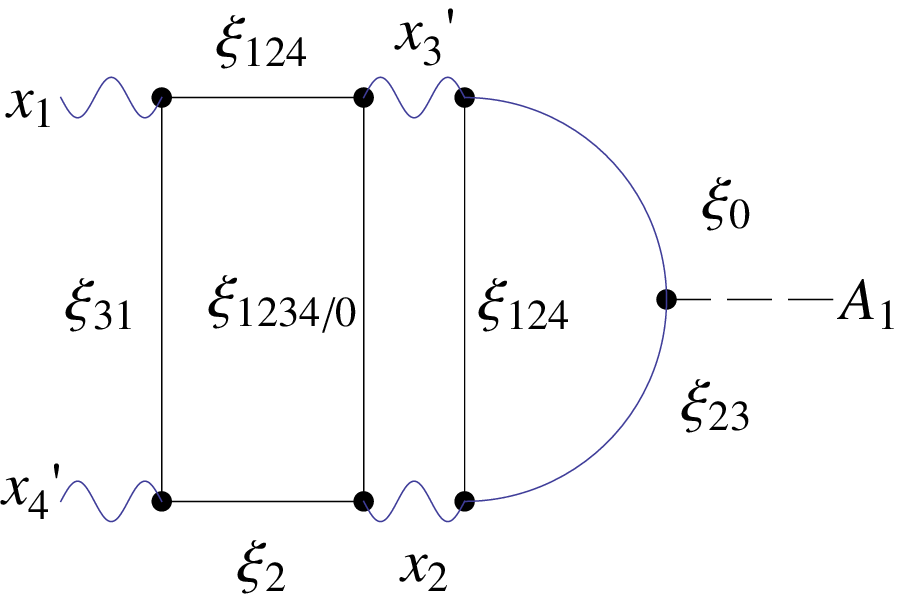}%23
\caption{The diagram via $A_1 x_3' x_2$ to $A_1 x_1 x_4'$.} 
\label{gb4b}
\end{center}
\end{minipage}
\hfill
\begin{minipage}[b]{0.47\linewidth}
\begin{center}
\includegraphics[width=6cm,angle=0,clip]{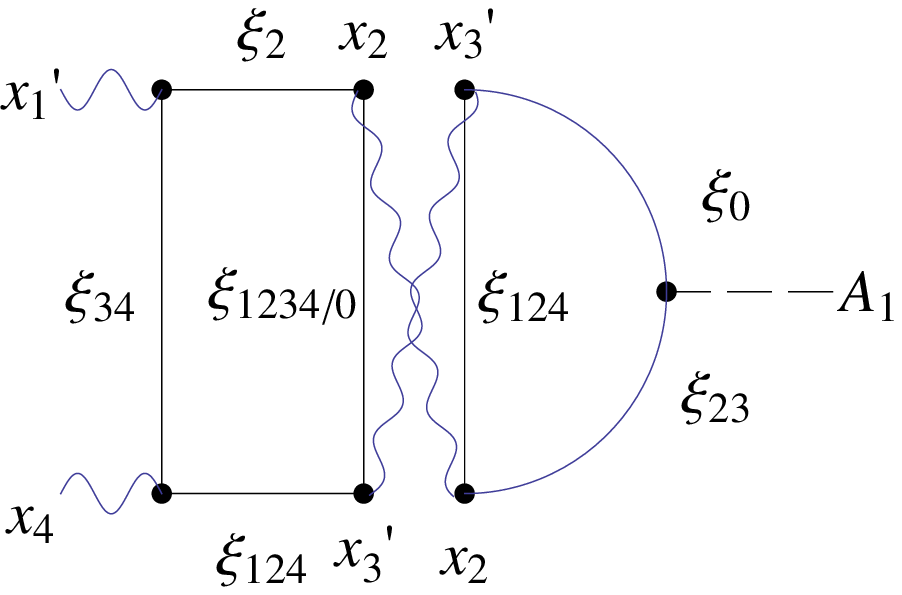}%24
\caption{The diagram via $A_1 x_3' x_2$ to $A_1 x_1' x_4$.}
\label{gb4dt}
\end{center}
\end{minipage}
\end{figure}

\begin{figure}
\begin{minipage}[b]{0.47\linewidth}
\begin{center}
\includegraphics[width=6cm,angle=0,clip]{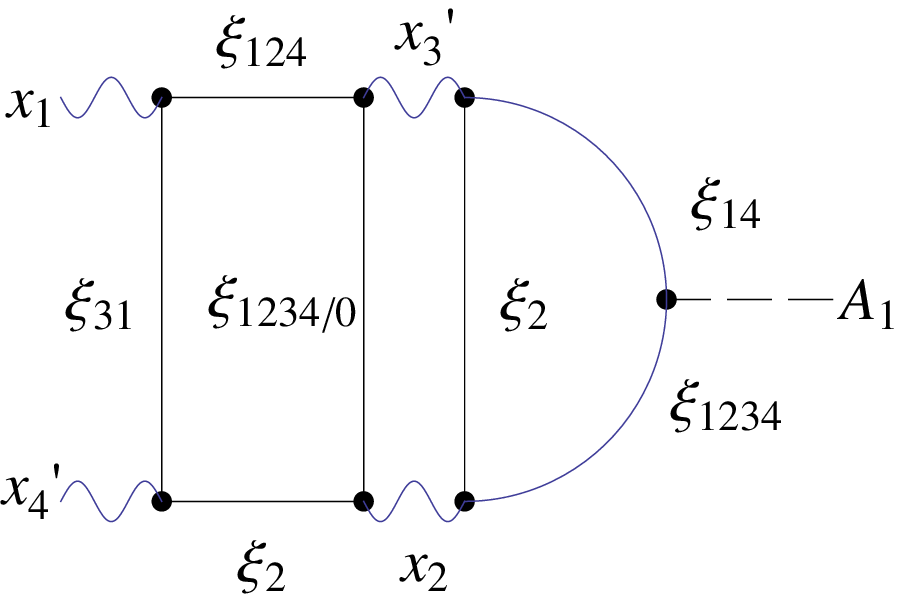}%23p
\caption{The diagram via $A_1 x_3' x_2$ to $A_1 x_1 x_4'$.} 
\label{gb4bp}
\end{center}
\end{minipage}
\hfill
\begin{minipage}[b]{0.47\linewidth}
\begin{center}
\includegraphics[width=6cm,angle=0,clip]{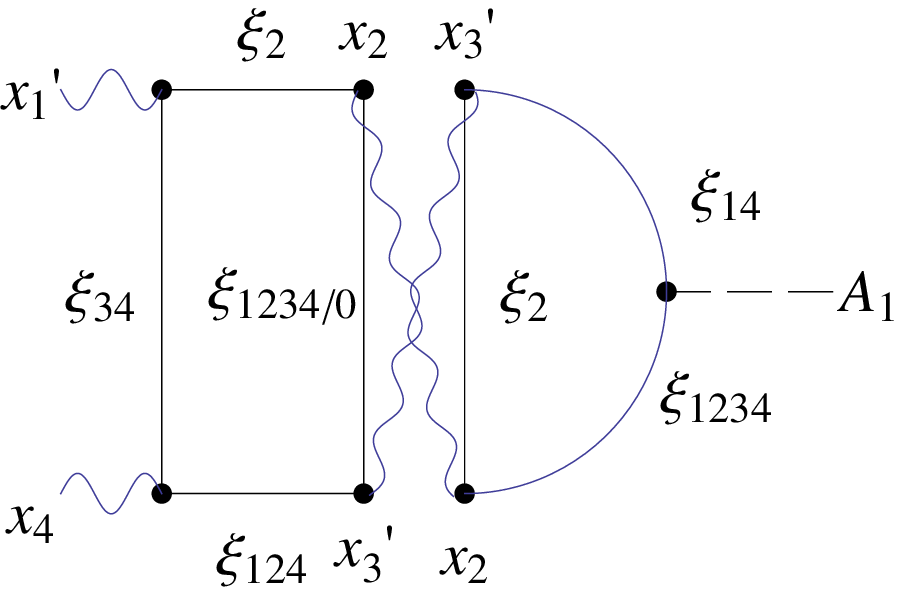}%24p
\caption{The diagram via $A_1 x_3' x_2$ to $A_1 x_1' x_4$.}
\label{gb4dpt}
\end{center}
\end{minipage}
\end{figure}

The diagrams for $A_2 x_1' x_3, A_2 x_1 x_3'$ and $A_3 x_2' x_1, A_3 x_2 x_1'$  are similar.

In the twisted diagrams, a mixing of $\xi_{1234}$ and $\xi_0$ is assumed, which appears after transformations of $G_{12}, G_{13}, G_{123}$ and $G_{132}$. The two vector particles which appear on the triangle diagram belong to different groups $E$ and $E'$, respectively. Such processes would contribute in the instanton and do not enter in decays of a Nambu-Goldstone boson.

A possible origin of large $\phi$ is a contribution of the twisted diagram, in which the 4th component of the quark that runs on the loop becomes a mixed state due to the operation of $G_{13}$ in the transition to the intermediate state.

%\newpage
\section{Discussion and Conclusion}

Whether \'E. Cartan's spinor matches the dynamics of QCD is not trivial. However, if physical theories are assembly of mathematical models or assembly of rules that connect elements of models and observables\cite{HM11}, there is a possibility that the triality symmetry plays a role in QCD. The possible role of the triality property of SO(8) group in the quark system was discussed in \cite{Sil95}. In my model, the triality selection rule in the electromagnetic interaction of leptons was incorporated, and no selection rule was assumed in the interaction of quarks.

  Photons emitted from matter made of quarks that belong to a triality sector different from that of electromagnetic probes,  will not be detected, and the matter will be assigned as a dark matter.  Dark matter search is done by the Xenon100 detector\cite{Xenon12}, and a direct detection of composite dark matter through lattice calculation of electromagnetic form fators and comparison to the data of Xenon100 was proposed\cite{LSD13}. The dark matter with masses less than 10TeV was excluded in this analysis.  However, the detection through electromagnetic probes in our world would be impossible due to the triality selection rule.

The dark matter search was done also in AMS-02 by measuring positron fraction in cosmic rays\cite{AMS02}. The signal that is detected in our electromagnetic probe would not originate from dark matter, but rather from pulsars, as a recent analysis suggests\cite{YBCGLZ13}.

A comparison of the decay width of  $\eta\to \gamma\gamma$ and $\eta'\to \gamma\gamma$ could be a place to study the effect of triality symmetry, including $\eta$ or $\eta'$ decay into intermediadte two vector particles in twisted diagrams. 
An investigation of the photon-gluon scattering contribution to the structure functions of deep inelastic scattering for unpolarized as well as polarized photons and gluons\cite{BINT91} could provide helpful information.  
The Primakoff production of $\eta$ and $\eta'$ in the Coulomb field of a nucleon also shows enhancement as compared to the production of $\pi$ mesons\cite{BDGHLL74a,BDGHLL74b,PrimEx11,KM11}.

In \cite{SF12b}, I considered three massless neutrinos in different triality sectors interacting with each other and produced one heavy and two degenerate light neutrinos. $\nu_e, \nu_\mu$ and $\nu_\tau$ have their lepton partners. I expect $e, \mu$ and $\tau$ are sensitive to flavors, but blind to the triality of neutrinos, quarks and gluons, and that they are sensitive to the triality of electromagnetic waves.  If electromagnetic waves from different triality sectors cannot be detected by electromagnetic probes in our world, we can understand the presence of dark matter. 

In the rescattering or twisted diagrams, $\bar q q$ state that decay into two vector particles appears. 
Brodsky and Shrock \cite{BS08a,BS08b} discuss problems in the expectation value of $\bar q q$ in QCD, which gives too large cosmological constant $\Omega_\Lambda$, and claimed that $\Omega_\Lambda$ has the spacial support within hadrons.  The recent review \cite{ClRo13} explains the region of matter distribution reachable with terrestrial facilities. Whether the spacial support and the region where the trialty sector agrees with that of our electromagnetic probes could match is under investigation.

\vskip 0.5 true cm
\leftline{\bf Acknowledgement}

The author thanks Craig Roberts for sending helpful references and Stan Brodsky for helpful information and comments. 

\newpage
%\vskip 1 true cm
\noindent{\large{\bf Appendix: Conjecture on the struture of the vacuum of universe}}

In order to obtain physical quantities like decay width of $\pi, \eta$ and $\eta'$ mesons, it is necessary to regularize Feynmann integral. L\"uscher\cite{Lu88} started from the space-time lattice
\[
\Lambda=\{x\,\epsilon\, {\bf Z}^4| -L/2<x_\mu\leq L/2, \mu=1,2,3,4\}
\]
and link variables $U(x,\mu)$ defined, when $x\,\epsilon\, \Lambda$ and $x+\hat\mu\,\epsilon\,\Lambda$.
The gauge transformation $g(x)$ of the gauge group $\mathcal G$ acts on the gauge field $\mathcal U$,
 and the compact Lie group $\mathcal G$ acts in a differential manner on the field manifold $\mathcal F$,
\[
g\,\epsilon\, {\mathcal G}, \quad U\,\epsilon\, {\mathcal F}\to g\cdot U=U^g\,\epsilon\, {\mathcal F}
\]
The orbit manifold ${\mathcal M}={\mathcal F}/{\mathcal G}$ is a differential manifold and with some measure on $\mathcal F$ defined as $d\mu(U)$ and some integrable function defined as $f(U)$, one introduces notations as follows.

We define basis of the Lie algebra ${\mathcal L}_G$ of $\mathcal G$ as $T^a, a=1,\cdots,d_{\mathcal G}$, general element of ${\mathcal L}_G$ as $X=X_a T^a$ and for any differential function $F(U)$ on ${\mathcal F}$
\[
\delta_X F(U)=X_a\left\{\frac{\partial}{\partial Y_a}F(e^{-Y}\cdot U)\right\}_{Y=0},
\]  
is defined.

For any subset of $\mathcal F$, defined as $\mathcal N$, a set
\[
[{\mathcal N}]=\left\{U\,\epsilon\, {\mathcal F}|g\cdot U\,\epsilon \,{\mathcal N}\quad {\rm for\quad some\quad g}\,\epsilon\, {\mathcal G}\right\}
\]
is defined as a union of all gauge orbits passing through $\mathcal N$.

It was shown in \cite{Lu88} that, by using invariant measure $dg$ on $\mathcal G$, 

\[
\chi(U)=\left\{\begin{array}{cc} 1& {\rm if}\, U \epsilon \, \mathcal N,\\
                   0 & {\rm otherwise}  \end{array}\right.               
\]
and the linear operator
\[
L(U)\cdot X=\delta_X F(U)\quad{\rm for\, all}\, X\, \epsilon {\mathcal L}_{\mathcal G},
\]
one can consider for a set $h\, \epsilon\, \mathcal G$ and $g=e^{-X}h$, a gauge fixing function
\[
F(g\cdot U)=L(h\cdot U)\cdot X+O(X^2).
\]

For any function $f(U)$ supported in $\mathcal N$, the integral
\[
\int_{\mathcal F}d\mu(U)f(U)\sim \int_{\mathcal F}d\mu(U)f(U)\int_{\mathcal G} dg\chi(g\cdot U)det L(g\cdot U)\delta(F(g\cdot U))
\]
reduces in the neighbourhood of $h$ to
\[
K\int_{\mathcal F}d\mu(U)f(U)  det\, L(U) \delta(F(U))
\]
where $K$ is a constant independent of $f$.

We consider the vacuum near $\mathcal N$ of our universe, and the universe transformed by $G_{23},  G_{12}, G_{13}, G_{123}$ and $G_{132}$.

%\newpage
\vskip 0.5 true cm


\begin{thebibliography}{99}
\bibitem{HM11} Hawking, S. and Mlodinow, L. {\it The Grand Design}, translated by Sato,K. , Kohdansha-pub. (2011).
\bibitem{GL84} Gasser, J. and Leutwyler,H. :Chiral Perturbation Theory: Expansions in the Mass of the Strange Quark, 
\NPB{\bf 250}(1984), 465.
\bibitem{WZ71} Wess, J. and Zumino, B.: Consequences of Anomalous Ward Identities, \PLB{\bf 37}(1971),95. 
\bibitem{Witten83} Witten,E. :Global Aspects of Current Algebra, \NPB{\bf 223}(1983),422.
\bibitem{KL00} Keiser, R. and Leutwyler, H. : Large $\it N_c$ in chiral perturbation theory, Eur. Phys. J. {\bf C17}(2000), 623, arXiv: 0007101[hep-ph].

\bibitem{BW01} Borasoy, B. and Wetzel, S.: U(3) chiral perturbation theory with infrared regularization, \PRD{\bf 63}(2001), 074019.
\bibitem{BB01} Beisert, N. and Borasoy, B.  : $\eta-\eta'$ mixing in U(3) chiral perturbation theory, Eur. Phys. J. {\bf  A11} (2001), 329: arXiv: 0107175 v1[hep-ph]

\bibitem{BN03} Borasoy, B. and Nissler, R. : Two-photon decays of $\pi^0,\eta$ and $\eta'$, Eur. Phys. J. {\bf 19} (2004), 367, arXiv: 0309011v2 [hep-ph]
\bibitem{KK06} Kekez, D. and Kabucar,D. :  $\eta$ and $\eta'$ mesons and dimension 2 gluon condensates $\langle A^2\rangle$, \PRD{\bf 73}(2006) 036002.
\bibitem{BCLRT07} Bhagwat,M.S., Chang,L. , Liu, Y-X, Roberts, C.D. and Tandy,P.C. : Flavour symmetry bleaking and meson masses, \PRC{\bf 76} (2007), 045203.
\bibitem{PS95} Peskin, M.E. and Schroeder,D.V.: {\it An Introduction to Quantum Field Theory},Perseus Books (1995).
\bibitem{AB69} Adler, S.L. and Bardeen,W.A. :Absence of Higher-Order Corrections in the Anomalous Axial-Vector   Divergence Equation, Phys. Rev.{\bf 182}, 1517 (1969).
\bibitem{AI89} Ansel'm,A.A. and Iogansen,A.A.: Radiative correction to the axial anomaly, JETP Lett. {\bf 49},214(1989).

\bibitem{Ioffe06} Ioffe,B.L.: Axial anomary: the modern status, arXiv:0611026[hep-ph].
\bibitem{Ioffe08} Ioffe,B.L.:Axial anomaly in quantum electro- and chromodynamics and the structure of the vacuum in quantum chromodynamics, Usp.Fiz.Nauk 178:647(2008), arXiv:0809.0212[hep-ph]
\bibitem{PDG12} Beringer,J. et al (Particle Data Group): Review of Particle Physics, \PRD{\bf 86}(2012), 010001.
\bibitem{DHL85} Donoghue, J.L., Holstein, B.R. and Lin, Y.-C.: Chiral Loops in $\pi^0, \eta^0\to\gamma\gamma$ and $\eta-\eta'$ Mixing, \PRL{\bf 55} (1985) 2766.
\bibitem{Shore01} Shore, G.M. : $\eta'(\eta)\to \gamma\gamma$: A Tale of Two Anomalies, Phys. Scripta T99 (2002), 84 , arXiv:[hep-ph/011165v1]
\bibitem{EF05} Escribano, R. and Fr\`ere, J-M. : Study of the $\eta-\eta'$ system in
two mixing angle scheme. JHEP 0506:029,2005: arXiv:0501072 v2[hep-ph]


\bibitem{MOU13} Michael, C., Ottnad, K. and Urbach, C. : $\eta$ and $\eta'$ mixing from Lattice QCD, arXiv:1310.1207v2[hep-lat].

\bibitem{Cartan66}  Cartan,\'E. {\it The theory of Spinors},p.118, Dover, New York (1966).
\bibitem{SF09} Furui, S.: Chiral Symmetry and BRST Symmetry Breaking, Quaternion Reality and Lattice Simulation, {\it Strong Coupling Gauge Theory in LHC Era}, p.398-400, World Scientific, Singapore (2011). 
\bibitem{SF10}  Furui, S.: Domain Wall Fermion Lattice Simulation in Quaternion Basis,  {\it The IX international Conference on Quark Confinement and the Hadron Spectrum-QCHS IX}, ed by Llanes-Estrada and Pela\'ez, AIP Conference Proceedigs 1343, p.533(2011), arXiv:0912.5397[hep-lat]
\bibitem{SF11} Furui,S.:Fermion flavors in quaternion basis and infrared QCD, Few Body Syst. {\bf 52}(2012), 171.
\bibitem{SF12a} Furui,S.: The magnetic mass of transverse gluon, the B-meson weak decay vertex and the triality symmetry of octonion, Few Body Syst.{\bf 53}(2012), 343.

\bibitem{SF12b} Furui,S.:The flavor symmetry in the standard model and the triality symmetry, Int. J. Mod. Phys. {\bf A27}(2012), 1250158.
\bibitem{SF13} Furui,S.:Axial anomaly and triality symmetry of octonion, Few Body Syst. {\bf 54}(2013), 2097, arXiv:1301.2095[hep-ph].


\bibitem{BINT91} Bass,S.D., Ioffe,B.L., Nikolaev, N.N. and Thomas,A.W.: On the Infrared Contribution to the Photon-Gluon Scattering and the Proton Spin Content, J. Moscow. Phys. Soc. {\bf 1}(1991), 317.


\bibitem{Sil95} Silagadze,Z.K.:SO(8) Colour as possible origin of generations,Yad.Fiz. 58 N8:1513-1517 (1995), arXiv:9411381[hep-ph].

\bibitem{Xenon12} Aprile,E. et al (XENON100 Collaboration): The XENON100 Dark Matter Experiment, Astropart. Phys. 35(2012), 573, arXiv:1107.2155[astro-ph.IM].

\bibitem{LSD13} Appelquist,T. et al (LSD Collaboration): Lattice calculation of composite dark matter form factors, arXiv:1201.1693[hep-ph].
\bibitem{AMS02} Aguilar, M. et al.:, First Results from the Alpha Magnetic Spectrometer on the International Space Station: Precision Measurement of thePositron Fraction in Primary Cosmic Rays of 0.5-350 GeV, \PRL{\bf 110} (2013), 141102.
\bibitem{YBCGLZ13} Yuan,Q. et al.: Implication of the AMS-02 positron fraction in cosmic rays, arXiv:1304.1482[astro-ph.HE]

\bibitem{BDGHLL74a}  Browman,A. et al. : Radiative Width of the $\eta$ Meson, \PRL{\bf 32}(1974), 1067.
\bibitem{BDGHLL74b} Browman,A. et al. : Decay Width of the Neutral $\pi$ Meson, \PRL{\bf 33}(1974), 1400.
\bibitem{PrimEx11} Latin, I. et al., (PrimEx Collaboration): New Measurement of the $\pi^0$ Radiative Decay Width, \PRL{\bf 106} (2011), 162303.
\bibitem{KM11} Kaskulov, M.M. and Mosel, U. : Primakoff production of $\pi^0,\eta$ and $\eta'$ in the Coulomb field of a nucleus, \PRC{\bf 84} (2011),065206, arXiv:1103.2097v2[nucl-th].
%\bibitem{BS08} Brodsky,S.J. and Shrock,R.: Maximum wavelength of confined quarks and properties of quatum chromodynamics, \PLB{\bf 666} (2008), 95, arXiv:0806.1535[hep-ph].
\bibitem{BS08a} Brodsky,S.J. and Shrock,R.: On Condensates in Strongly coupled Gauge Theories, Proc. Nat. Acad. Sci {\bf 108} (2011), 45-50, arXiv:0803.2541[hep-th]. 
\bibitem{BS08b} Brodsky,S.J. and Shrock,R.: Standard-Model Condensates and the Cosmological Constant, Proc. Nat. Acad. Sci {\bf 108} (2011), 45-50, arXiv:0803.2554[hep-th].
\bibitem{ClRo13} Clo\"et, I.C. and Roberts, C.D.: Explanation and Prediction of Observables using Continuum Strong QCD, arXiv:1310:2651[nucl-th]
\bibitem{Lu88} L\"uscher,M. :Selected Topics in Lattice Field Theory, E. Brezin and J. Zinn-Justin, eds., Les Houches, Session XLIX, 1988, Fields, Strings and Critical Phenomena, Elsevier (1989).
\end{thebibliography}
\end{document}